\documentclass[10pt]{iopart}

\usepackage{iopams}
\usepackage{graphicx}
\begin{document}

\title [Reversible control of interface-water induced carrier density in graphene] {Reversible control of interface-water induced carrier density in graphene-on-SiO$_2$ by thermal cycling under gate-voltage}.

\author{Anil Kumar Singh \& Anjan Kumar Gupta}

\address{Department of Physics, Indian Institute of Technology Kanpur, Kanpur 208016, India}
\ead{anjankg@iitk.ac.in}
\vspace{10pt}
\begin{indented}
\item[]{\today}
\end{indented}

\begin{abstract}
A reversible handle on graphene carrier density, other than the gate electric field, is desirable for memory applications of graphene. Our experiments show that the commonly observed carrier density hysteresis in graphene on SiO$_2$ due to interface water/oxygen vanishes at temperatures below 250 K. The state of these species, which affects the graphene carrier density, at low temperatures can be reversibly controlled by thermal cycling to room temperature at different gate voltages. Further, devices prepared with relatively dry interface, and thus showing negligible hysteresis at room temperature, show a marked increase in hysteresis on heating above room temperature. Thus thermal-cycling, to high temperatures and under gate-voltage, provides a reversible handle on carrier density. These results are discussed in terms of temperature and interface-water density dependence of redox-reaction kinetics.
\end{abstract}

%
\vspace{2pc}

\maketitle
%
%

\section{Introduction}

Graphene has shown great application potential in sensing and memory devices \cite{Ohno-2010,Kumar-2013,Zhan-2014,Park-2016} due to the sensitivity of its electronic properties to the environment and underlying substrates. Existence of hysteresis in resistance with gate voltage ($V_g$) in field electron transistor (FET) based on 2D materials \cite{Choi-2013,Late-2012,Tan-2014,Zhang-2016,Cadore-2016} raises the possibility of a reversible control of carrier density by means other than $V_g$. Hysteresis has been seen in majority of the graphene devices prepared under varying conditions on various substrates \cite{Tan-2014,Cadore-2016,Wang-2010,Veligura-2011,Mohrmann-2015,Bharadwaj-2016}. Two types of hystereses, negative and positive, have been reported in graphene FETs \cite{Wang-2010,Veligura-2011}. The negative (positive) hysteresis occurs due to enhancement (reduction) of electric field near graphene leading to a shift in $V_g$ corresponding to charge-neutrality point, i.e. $V_{cnp}$, towards positive (negative) values for negative $V_g$. An example of negative hysteresis is the electrolyte and ferroelectric gated graphene devices \cite{Wang-2010,Zheng-2010,Rogers-2017}.

In order to understand the positive hysteresis in graphene on SiO$_2$, several research groups have studied these devices under various conditions using transport \cite{Wang-2010,Veligura-2011,Levesque-2011} as well as Raman \cite{Xu-2012} studies. The most commonly believed hysteresis mechanism \cite{Xu-2012,Veligura-2011,Levesque-2011} in graphene on SiO$_2$ involves charge transfer between graphene and redox species, derived from H$_2$O and O$_2$, due to the difference in electrochemical potential of the redox reaction and graphene. This also leads to doping, predominantly p-type, of graphene in the absence of gate electric field. The presence of H$_2$O and O$_2$ at graphene/SiO$_2$ interface is inevitable unless the devices are vacuum annealed at high temperatures. However, post-annealing exposure to ambient conditions leads to slow intercalation of these species \cite{Lee-2014}. Further, since graphene Fermi energy can be controlled by the gate electric-field one can get reversible, though slow \cite{Xu-2012, Liu-2011}, transfer of charge between graphene and redox species. The time scale of charge-transfer depends on the redox reaction kinetics described by Marcus-Gerischer theory \cite{Levesque-2011, Xu-2012}. Although several reports on hysteresis due to charge transfer between graphene and interface-states exist but, by contrast, not much attention has been given to the control of CNP using this charge transfer process.

\begin{figure}
\centering
\includegraphics[width=8.0cm]{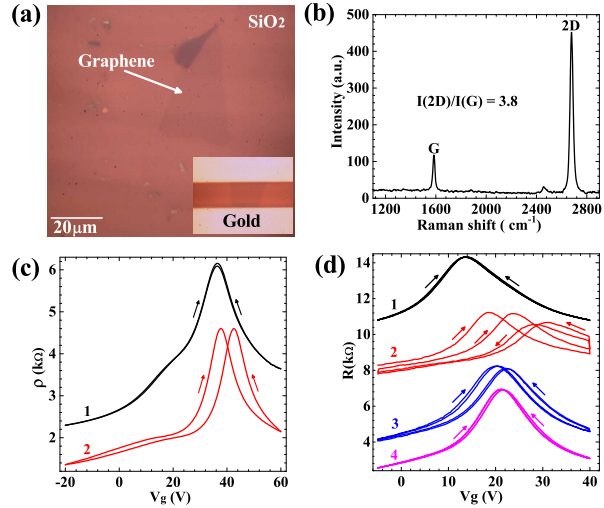}
\caption{ (a) Optical image of graphene on SiO$_2$; the inset shows the same graphene after depositing Cr/Au contacts. (b) shows the Raman Spectra of the graphene using 532nm and 0.50 mW laser. (c) resistivity vs V$_g$ of two different graphene devices; curve-1 (1.5 k$\Omega$ upward shift for clarity) is for a dry-interface device while curve-2 is for a PSC device with wet-interface. (d) shows the effect of water exposure on a non-hysteretic device depicted by curve-1 (8 k$\Omega$ shift) in which the hysteresis appears, as per curve-2 (5.5 k$\Omega$ shift) after exposure to a drop of water. The hysteresis weakens in curve-3 (1.5 k$\Omega$ shift) taken after 5 hrs in vacuum and disappears in curve-4, which is taken after 11 hrs in vacuum.}
\label{fig:Fig1}
\end{figure}

Lithography free graphene devices \cite{Tien-2016} show high sensitivity for both gas and liquid environments due to lack of disorder and charge trapping by polymer residue. It is also known that graphene on SiO$_2$ faces more interface traps and disorder as compared to suspended graphene \cite{Mayorov-2012} or graphene on hexagonal boran nitride. Therefore for better performing graphene-FETs SiO$_2$ is best avoided; however, with a reversible handle on the state of interface traps, such devices can be useful for data storage \cite{Imam-2011}. Doping in graphene on SiO$_2$ has also been controlled by influencing charge-traps with electromagnetic radiation of different wavelengths \cite{Ho-2015, Pavel-2017}.

In this paper, we present a study on graphene carrier-density control by using the hysteretic state in lithography-free graphene-on-SiO$_2$ devices. On cooling these devices the hysteresis disappears below 250K fixing $V_{cnp}$ and the frozen-state of the interface-traps determining $V_{cnp}$ value then depends on $V_g$. This gives a reversible control on $V_{cnp}$. Further, in devices with dry interface and negligible hysteresis at room temperature, significant hysteresis appears on heating above room temperature. In this case, cooling from above room temperature under $V_g$ provides a reversible handle on $V_{cnp}$. A qualitative understanding of these results is discussed in terms of redox-reaction kinetics determined by the diffusion of species and redox-reaction barrier.

\section{Experimental details}

n-doped Si substrates with 300 nm thick thermal SiO$_2$ were first cleaned with acetone and iso-propyl alcohol (IPA). The substrates were then either oxygen plasma cleaned (OPC) with 50 W power for 10 min or piranha solution cleaned (PSC) in fresh piranha solution for 10 min. The PSC substrates were also dipped/cleaned in deionized water followed by flushing with dry N$_2$ gas. The water exposure in OPC substrates was optionally done for experimenting with water trapped at graphene-SiO$_2$ interface. Graphene was mechanically exfoliated from Kish graphite on these substrates using an adhesive tape. For relatively dry-interface devices, graphene was stamped on IPA cleaned substrate after heating it to 100 $^\circ$C. Raman spectrum in figure\ref{fig:Fig1}(b) shows the characteristic monolayer graphene features i.e. G and 2D bands with I(2D)/I(G)= 3.8 \cite{Ferrari-2006}. Mechanical wire-masking \cite{Singh-2016} was used to make two probe Cr(10 nm)/Au(50 nm) contacts on graphene, see figure\ref{fig:Fig1}(a) inset. Two probe resistance down to liquid nitrogen temperatures was measured in a homemade vacuum cryostat with a heater for temperature control. A 10 k$\Omega$ resistance was used with the gate voltage supply, which was controlled by a data acquisition card using a LabView program. The cryostat was pumped using a turbo molecular pump capable of better than $10^{-4}$ mbar pressure. When the cryostat is dipped in liquid nitrogen for cooling, the pressure is expected to be much lower than this.

\section{Experimental results}

\subsection{Hysteresis due to water at graphene-SiO$_2$ interface}
Two probe transport measurements were performed on several lithography-free graphene devices. Both types, i.e. OPC and PSC, of devices show significant p-doping due to silanol group density on SiO$_2$, which makes the surface hydrophilic \cite{Nagashio-2011}. An increase in doping magnitude with OPC and PSC process times was also observed. V$_g$ dependence of resistance on dry-interface devices shows negligible hysteresis in vacuum, see curve-1 in figure\ref{fig:Fig1}(c). On the other hand, devices with interface water show significant hysteresis at room temperature and even after 48 hrs in vacuum, see curve-2 of figure\ref{fig:Fig1}(c). There was some reduction in hysteresis during initial pumping due to removal of water on top of graphene. For further confirming that the positive hysteresis in our devices is predominantly from the interface water, we expose a dry-interface device to a drop of water for a few minutes to look at its effect on hysteresis. Initially the device showed small p-doping and no hysteresis, see figure\ref{fig:Fig1}(d), and it became hysteretic on water exposure; but the hysteresis disappeared after 11 hrs of pumping at room temperature.

\begin{figure}
\centering
\includegraphics[width=8.0cm]{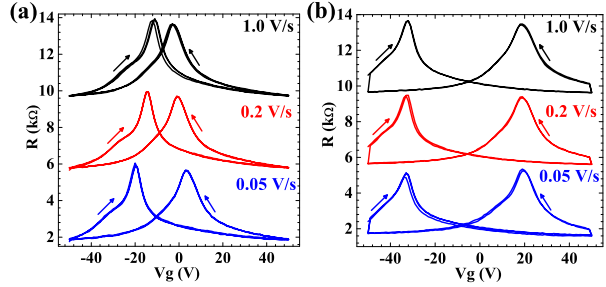}
\caption{R Vs $V_g$ of a four months old graphene device showing (a) the effect of sweep rate with no charging time and (b) effect of sweep rate with 1 hour charging time. The curves have been shifted uniformly by 4 k$\Omega$ for clarity.}
\label{fig:Fig2}
\end{figure}

Hysteresis was found to depend on V$_g$ range, sweep rate \cite{Veligura-2011, Mohrmann-2015} and the waiting time at the extreme $V_g$ values i.e. charging time. Figure \ref{fig:Fig2}(a) shows the effect of sweep rate on hysteresis in a device which has strong positive hysteresis due to interface water. The difference between the $V_{cnp}$ values, i.e. $\Delta V_{cnp}$, of the forward (-ve to +ve) and backward (+ve to -ve) gate-sweep is 8.8 V, 13.7 V and 23.4 V for sweep rates of 1V/s, 0.2V/s and 0.05V/s, respectively. For 60 min charging time $\Delta V_{cnp}$ = 51.5 V and, it is independent of the sweep rate as seen in figure \ref{fig:Fig2}(b). The detailed relaxation behavior in graphene FETs exhibits multiple time scales as reported in literature \cite{Lee-2013}. We also find, specifically at extreme gate voltages, multiple relaxation times including a very fast time scale together with a distribution of slow-ones giving a stretched exponential behavior with exponent close to 0.5. Thus even after an hour of wait an extremely slow relaxation was still seen.

\begin{figure}
\centering
\includegraphics[width=8.0cm]{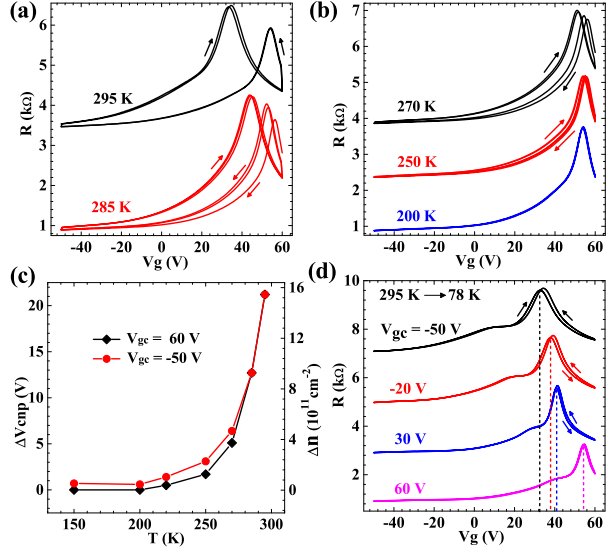}
\caption{(a) and (b) show the effect of cooling on R vs V$_g$ hysteresis at V$_{gc}$ = 60 V on a PSC device with wet-interface. Uniform vertical shifts of 2.5 k$\Omega$ and 1.5 k$\Omega$ have been used in (a) and (b), respectively. (c) shows temperature dependence of $\Delta V_{cnp}$ and the corresponding change in the charge density in electron-charge units, $\Delta n=\kappa\Delta V_{cnp}\epsilon_0/ed$, in the interface at two different V$_{gc}$ values. Here $\kappa=4$, $\epsilon_{0}$ is the free space permeability, $d=300$ nm and $e$ is the electron charge. (d) shows effect of V$_{gc}$ on the CNP (2 k$\Omega$ shift). A 30 min charging time was used in acquiring these data.}
\label{fig:Fig3}
\end{figure}

\subsection{Hysteresis evolution with cooling}

From above studies we believe that after keeping the devices in high vacuum for long time the redox reactions \cite{Levesque-2011,Xu-2012} due to the trapped H$_2$O/O$_2$, at the interface, are predominantly responsible for positive hysteresis in our devices. Removal of these interface species has been demonstrated by vacuum annealing \cite{Romero-2008} for better behaved FETs. On the other hand, the hysteresis can help in getting a reversible handle on graphene carrier density, other than the gate electric field. Water is expected to freeze below 200 K at any pressure while above this temperature it may exist as gas, liquid or solid depending on pressure. Thus if the positive hysteresis here is due to trapped water it should respond to cooling. Figure \ref{fig:Fig3}(a)-(c) show that the hysteresis reduces sharply from 295 to 250 K, and eventually disappears. The cooling was done under an applied gate voltage, namely $V_{gc}$, of 60 V and thus we see a large hole doping. In these measurements, the sample was warmed to 295 K and kept at $V_g$ = 60 V for 30 min before cooling it down to required temperatures. A charging time of 30 min was also used to get maximum, and sweep rate independent, hysteresis.

Since -ve $V_g$ gives electron doping we expect to see a non-hysteretic behavior with less p-, or even n-, doping for negative $V_{gc}$. Figure \ref{fig:Fig3}(d) shows the $V_{gc}$ dependence of R Vs $V_g$ at 78 K. Again, each R Vs $V_g$ scan was obtained at 78 K after keeping the sample at 295 K at desired $V_{gc}$ for 30 min and then cooling down to 78 K at the same $V_{gc}$. We see negligible hysteresis at 78 K but $V_{cnp}$ changes significantly with $V_{gc}$ value. As seen in figure \ref{fig:Fig3} (d), there is a systematic trend of this little unfrozen hysteresis, which is seen down to 5 K in some of the samples. As $V_{gc}$ is changed from -ve to +ve values this residual hysteresis vanishes and the respective peaks at $V_{cnp}$ become sharper. The later could arise from either an enhancement in mobility or from an increment in the gate capacitance with $V_{gc}$. The mobility enhancement could occur due to reduction in the density of ionized traps at the interface while the capacitance depends on the dielectric properties of either SiO$_2$ or the interface species on $V_{gc}$. This remains to be investigated.

\begin{figure}
\centering
\includegraphics[width=8.0cm]{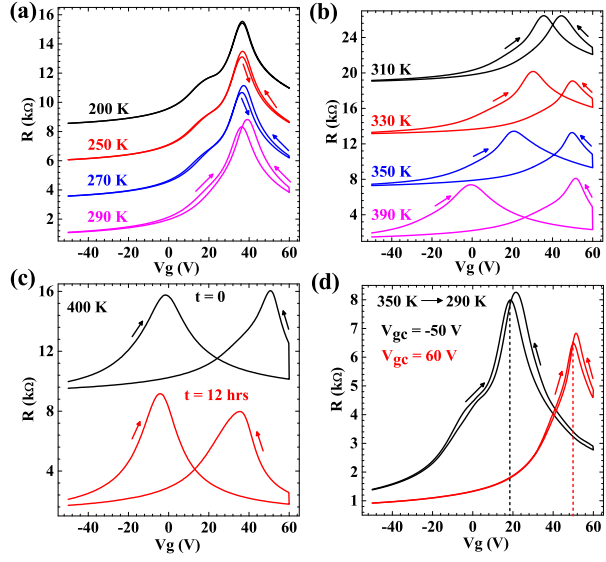}
\caption{Effect of cooling (a) and heating (b) on R vs V$_g$ hysteresis on a dry-interface device. Uniform vertical shifts of 2.5 k$\Omega$ and 6 k$\Omega$ have been used in (a) and (b), respectively. (c) shows the effect of 12 hrs waiting time at 400 K with 8.0 k$\Omega$ offset. (d) shows the effect of V$_{gc}$ on CNP. These data were taken with a charging time of 30 min and 0.2 V/s sweep rate.}
\label{fig:Fig4}
\end{figure}

Figure \ref{fig:Fig4} shows temperature and $V_{gc}$ dependence of R Vs $V_g$ of a device with dry-interface, and thus with very small hysteresis, at room temperature in vacuum. As expected, the hysteresis diminishes with cooling down to 200 K; however, it increases with heating. We notice that the hysteresis increase is more due to shift of $V_{cnp}$ of forward scan towards negative $V_g$ values than the shift of reverses-scan $V_{cnp}$ towards higher $V_g$ values. Also when the sample is held at 400 K for 12 hours the hysteresis reduces, see figure \ref{fig:Fig4}(c), due to slow desorption of interface species. In detail, after 12 hrs both the resistance peaks shift towards negative $V_g$ and the right peak shifts more leading to a reduction in $\Delta V_{cnp}$. All these different temperature curves were acquired while heating at $V_g=0$ V and with a charging time of 30 min. Figure \ref{fig:Fig4}(d) shows the effect of $V_{gc}$ on cooling from 350 K to 290 K demonstrating the controllability of $V_{cnp}$ at 290 K in devices showing very small hysteresis at 290 K. Other detailed features, such as a secondary hump in figure \ref{fig:Fig4} (a) and (d) or resistance value at respective $V_{cnp}$, and their evolution with temperature and $V_{gc}$ are yet to be understood.

\section{Discussion}

The direction of electrochemical reaction \cite{Levesque-2011,Xu-2012,Chakrapani-2007}, responsible for doping and hysteresis, is dictated by the direction of electron transfer between graphene and redox species. Thus when graphene chemical potential ($\mu_g$), which is -4.6 eV (relative to the vacuum level) for charge neutral graphene, is not equal to the electrochemical potential ($\mu_r$) of the redox species, electron exchange between the two will take place. For instance H$_2$O/O$_2$ redox species in equilibrium in ambient conditions have $\mu_r$ value of -5.3 eV \cite{Levesque-2011,Xu-2012}. The $\mu_r$ value, for the same species trapped at the interface between SiO$_2$ and graphene, need not be the same. Since we see hole doping in most devices, it is reasonable to assume that $\mu_r<$ -4.6 eV leading to electron transfer from graphene to redox species. This also requires the graphene work function to remain pinned to its bands so that any Fermi energy change leads to an equal magnitude change in work function; this is seen to be the case from recent experiments \cite{Yuan-2015}. From the typical hole density (deduced from $V_{cnp}$ values) of $\sim 10^{12}$ cm$^{-2}$ in our devices, the change in $\mu_g$ is about 0.15 eV implying that if graphene and redox species are in equilibrium, $\mu_r \approx$ -4.75 eV when $V_g=0$. Here, we have ignored the direct effect of gate electric field on the redox reactions at the interface.

A positive $V_g$ will increase $\mu_g$ and promote electron transfer from graphene to interface species for electrochemical reactions that happen slowly. Thus when one goes from negative (positive) $V_g$ to the positive (negative) extreme of the $V_g$, the gate induced increase (decrease) in $\mu_g$ leads to electron transfer to (from) the interface species from (to) graphene. This changes the charge-state of the interface species and reduces the magnitude of the gate electric-field felt by graphene. Thus, during the charging time, the carrier density in graphene progressively reduces, leading to an increase in resistance as seen clearly in all devices having large hysteresis. The extra charge thus created at interface gives rise to opposite sign carriers, via electrostatic interaction, in graphene. If one now sweeps $V_g$, faster than the reaction time-scale, negligible charge transfer happens between graphene and interface species. This leads to a carrier density change mainly due to change in gate electric-field. The $V_g$ value, history and temperature dependence of this reaction time scale have rather complex physics and need further work. From the sweep-rate independence after long charging time it seems that the reaction rate slows down significantly after the wait at the extreme $V_g$ values.

From above results it is clear that the freezing of water at 200 K is not critical for no-hysteresis but the density of water at the interface plays an important role. Higher temperatures would promote diffusion as well as electrochemical reaction of interface species by enhancing the activation over the energy-barriers for diffusion and reaction. The separation between required species will depend on their density and thus at low density the species need to diffuse through longer distances for reaction. Thus for low density of interface water higher temperatures will be required for significant reaction-rates and for hysteresis in dry-interface devices. The thermal cycling under gate-voltage gives a handle, other than $V_g$, on $V_{cnp}$; however, a much faster control, by way of controlling diffusion and reaction rates of interface species, is necessary for actual memory applications.

\section{Conclusions}

In summary, the carrier density hysteresis in graphene on SiO$_2$ devices in high-vacuum can be controlled by the density of water at the interface and one can get a negligible-hysteresis state at room temperature for low interface water density. Further, the presence of hysteresis at high temperatures allows a reversible control on the graphene carrier density. Finally, while the electrochemical cell memory \cite{ECM-2011} is being explored as an alternative to the flash memory, this work may lead to another memory device combining the field effect, used in flash, and electrochemistry.

\section*{Acknowledgement}

The authors acknowledge financial support from IIT Kanpur, CSIR and DST of the Government of India.

\end{document}